\documentclass[10pt,aps,prl,twocolumn,amssymb,floatfix,superscriptaddress]{revtex4}
\usepackage{amsmath,bm,epsfig,graphics,mathrsfs}

\newcommand{\tx}{\text}
\newcommand{\ti}{\textit}

\newcommand{\alp}{\alpha}

\newcommand{\Dlt}{\Delta}

\newcommand{\gm}{\gamma}

\newcommand{\sg}{\sigma}

\newcommand{\ssf}[1]{\tx{\tiny{#1}}}

\begin{document}
\title{Geometrically Enhanced
Quantum Oscillatory Signal and Nonzero Berry's Phase in Indium Arsenide Surface}
\author{Jian Sun}
\thanks{Corresponding author: sun-jian@jaist.ac.jp}
\affiliation{Computer, Electrical and Mathematical Sciences and
Engineering Division, King Abdullah University of Science and
Technology (KAUST), Thuwal 23955-6900, Saudi Arabia}
\affiliation{School of Materials Science, Japan Advanced Institute of
Science and Technology, Nomi 923-1211, Japan}
\author{Xuhui Wang}
\thanks{Corresponding author: xuhuiwangnl@gmail.com\\}
\affiliation{Physical Sciences and Engineering Division, King
Abdullah University of Science and Technology (KAUST), Thuwal
23955-6900, Saudi Arabia}
\author{Sadamichi Maekawa}
\affiliation{Advanced Science Research Center, Japan Atomic Energy
Agency, Tokai 319-1195, Japan}
\affiliation{CREST, Japan Science
and Technology Agency, Tokyo 102-0075, Japan}
\author{Aur\'{e}lien Manchon}
\affiliation{Physical Sciences and Engineering Division, King
Abdullah University of Science and Technology (KAUST), Thuwal
23955-6900, Saudi Arabia}
\author{J\"{u}rgen Kosel}
\affiliation{Computer, Electrical and Mathematical Sciences and
Engineering Division, King Abdullah University of Science and
Technology (KAUST), Thuwal 23955-6900, Saudi Arabia}
\date{\today}

\begin{abstract}
In a system accommodating both surface and bulk conduction channels, a
long-standing challenge is to extract weak Shubnikov-de Haas
oscillation signal in the surface from a large background stemming
from the bulk. Conventional methods to suppress the bulk
conduction often involve doping, an intrusive approach, to reduce
the bulk carrier density. Here we propose a geometric method, i.e.
attaching a metal shunt to the indium arsenide epilayer, to
redistribute current and thus enhance the
oscillation-to-background ratio. This allows us, for the first
time, to observe clear quantum oscillations and nonzero Berry's
phase at the surface of indium arsenide. We also identify the
existence of a Rashba type spin-orbit interaction, on the InAs
surface, with a large coupling constant $\alp\sim 1 ~
\tx{eV}~\tx{\AA}$. We anticipate wide applicability of this
non-intrusive architecture in similar systems such as topological
insulators.
\end{abstract}
\maketitle

Indium arsenide (InAs), an ordinary narrow-gap semiconductor with
high intrinsic mobility and considerable field effect, is a
promising material in electronic applications
\cite{ko-nature-2010}. Yet a surface state, with a sheet density
of $\sim 10^{12}~\tx{cm}^{-2}$ residing on the free surfaces of
InAs as due to band bending, has been known for decades
\cite{olsson-prl-1996}. The recent rise of topological insulators
(TIs) has enriched the pool of host materials with surface state
\cite{Kane-TI}. Dimensionality and energy dispersion of these TI
surface states are frequently studied using transport measurements
such as the Shubnikov-de Haas (SdH) oscillation
\cite{shoenberg-book}. Meanwhile, the quantum oscillations from
the free surfaces intrinsic to InAs was seldom reported. The
regular Hall-bar-type geometry measures the sample as a
homogeneous system and thus provides an averaged signal. The
intimate contact between the surface state and underlying bulk,
however, impedes the extraction of weak SdH signal of the former
from a large resistance background of the latter. In TIs, for
example, it is a challenging task to suppress the bulk carrier
density to enhance the surface dominated transport
\cite{analytis-natphys-2010}.

In this letter, we propose a novel architecture to remove this
long-standing obstacle. We use a metal-shunted Hall bar to enhance
the SdH signal in the InAs surface. The enhancement achieved here
is often called \textquotedblleft geometric \textquotedblright ,
i.e. modifying a conductor by changing its geometric shape, which
has been defined for the enhanced magnetoresistance in
semiconductors \cite{Allgaier-JAP, solin-science}. We apply this
architecture on InAs to extract efficiently the SdH signals in the
surface state. From the SdH Oscillations, we
further discover a nonzero Berry's phase as a
manifestation of the Rashba coupling driven by the inversion
asymmetry.


\begin{figure*}[t]
\centering
\includegraphics[scale=0.9]{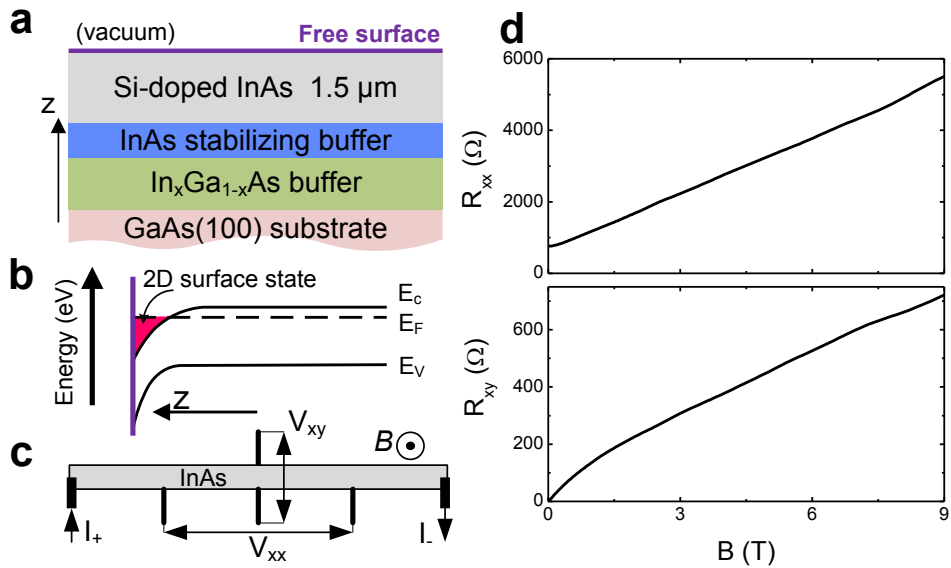}
\caption{N-doped InAs epitaxial sample (a) Layer diagram of metamorphic
buffered Si-doped InAs epitaxial sample. (b) Schematic of the formation of
2D surface state as the result of band bending. (c) Sketch of a regular Hall bar device for transport measurement. (d) Longitudinal magnetoresistance $R_{xx}$ and Hall resistance $R_{xy}$ measured in perpendicular magnetic fields at 10 K in the regular Hall bar device.}
\label{fig:fig1}
\end{figure*}

The experimental sample is a $1.5~\mu$m thick Si-doped InAs
epilayer on GaAs substrate (Fig.~\ref{fig:fig1}a, also see
Methods). The metamorphic buffer layer technique is employed to
absorb strain from lattice mismatch and prevent vertical
propagation of dislocations to maintain the quality of the top
InAs active layers \cite{meta_Lub}. It
also aligns the energy bands between GaAs and InAs, avoiding
possible formation of two-dimensional electron systems across the
heterojunction. We are confident to argue that any signals of 2D
nature shall originate from the InAs surface state (see
Fig.~\ref{fig:fig1}b).

As a control experiment, we first show the transport measurements
in a regular InAs Hall bar \ti{without} a metal shunt, as sketched
in Fig.~\ref{fig:fig1}c. The longitudinal magnetoresistance
$R_{xx}$ and Hall resistance $R_{xy}$ at perpendicular magnetic
fields are shown in Fig. \ref{fig:fig1}d. Two curves are smooth,
as the large background signals wash out the quantum oscillations.
The nonlinearity in $R_{xy}$ at low field implies the existence of
a parallel conducting channel, i.e. the surface state
\cite{Wieder-APL-1974, xiong-prb-2012}.


\begin{figure*}[t]
\centering
\includegraphics[scale=0.7]{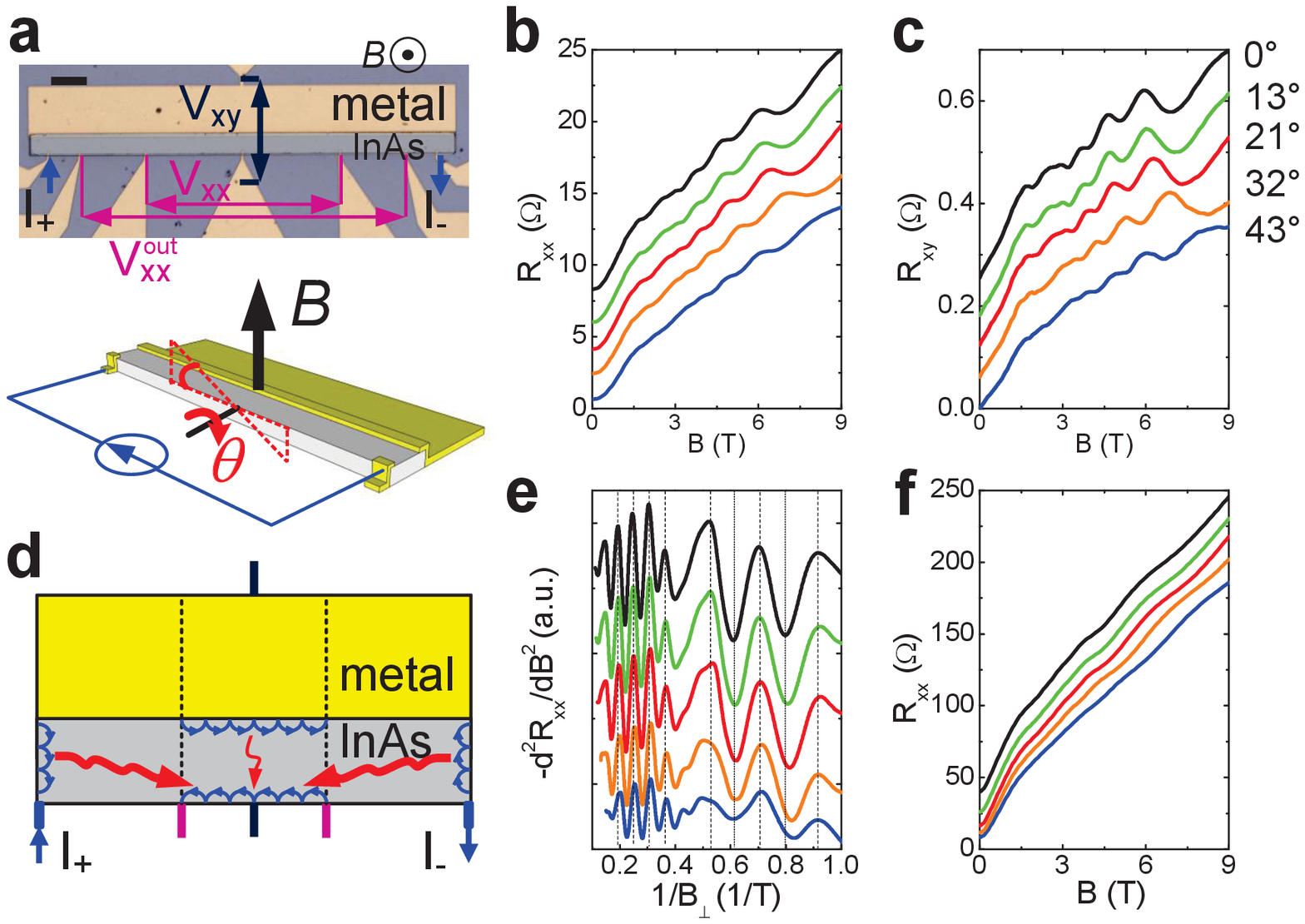}
\caption{Magnetotransport measurements with shunted Hall bar (a)
Optical image (top-view) of the shunted Hall bar. Scale bar is 100 $\mu$m.
The cartoon shows the measurement configuration in a magnetic field.
(b) $R_{xx}$ and {\bf (c)} $R_{xy}$ measured in titled magnetic fields of
angle $\theta$ at 10 K. Curves measured at different angles are
distinguished by colors and are shifted vertically for clarity.
(d) Sketch of the edge states in the shunted structure. The blue curves at the edges
represent the skipping orbits, i.e., the edges states. The wavy red curves with arrows represent the scattering that contributes to the SdH oscillations.
(e) $-d^{2}R_{xx}/dB^{2}$ of (b) as a function of $1/B_{\perp}$. The dash lines locate the extrema of the oscillations.
(f) $R_{xx}$ measured using the additional
outer electrodes V$_{xx}^{\tx{out}}$ at 10 K.}
\label{fig:fig2}
\end{figure*}

The shunted Hall bar, shown in Fig. \ref{fig:fig2}a, is prepared
via conventional lithography procedures (see Methods). At 10 K,
the $R_{xx}$ and $R_{xy}$ measured in the shunted bar are shown in
Fig. \ref{fig:fig2}b and c, respectively. In contrast to the
smooth curves observed in the regular Hall bar, both $R_{xx}$ and
$R_{xy}$ exhibit pronounced oscillation superimposed on a
suppressed background (see the values on the vertical axis), or,
an enhanced oscillation-to-background ratio has been achieved.

To uncover the origin of the oscillations, we scan the angular
dependence of the SdH oscillations by rotating the sample in a
magnetic field $B$ to different angles $\theta$ (Fig.
\ref{fig:fig2}a).  In Figs. \ref{fig:fig2}b and c, local minima
(maxima) show a systematic shift against $B$, whereas in Fig.
\ref{fig:fig2}e,
they are aligned against $1/B_{\perp}$ at all angles.
This clearly exhibits the 2D nature \cite{qu-science-2010,
analytis-natphys-2010} of the surface state in InAs.

Intriguing yet simple mechanisms are driving the enhancement. The
metal shunt redistributes the current injected into the InAs bar
as electrons favour the paths of the lowest resistance, according
to the Ohm's law. Finite element simulations reveal an
inhomogeneous current distribution in the InAs bar.
Most electrons flow into(out of) the metal shunt along the
$y$-direction near two current leads;
while a small portion of them traverses the InAs bar without
entering the shunt. The middle part of the InAs bar, i.e. between
two $V_{xx}$ electrodes, is effectively shorted by the metal. The
background signal measured locally at $V_{xx}$ is thus much
suppressed, again, according to Ohm's law.

The local current reduction seems to suggest a weakened
oscillatory signal too. The SdH oscillations in the surface layer,
however, originate from the scattering between edge states
localized at InAs bar boundaries
\cite{buttiker-prb-1988,vanwees-prb-1989}, demanding a global
picture on the entire conductor (Fig. \ref{fig:fig2}d). In our
geometry, as a current is injected under a perpendicular magnetic
field, abundant edge states are formed close to the left and right
edges of the InAs bar. In the 2D surface, the scattering events
also happen between edge states at the left (right) and lower
edges. Therefore, the SdH signal probed at the $V_{xx}$ contacts
is an average over contributions from three edges (Fig.
\ref{fig:fig2}d). The reduction of the SdH signal is hence less
significant than the suppression in the background. And the
oscillation-to-background signal ratio is enhanced. This mechanism
relies on the geometry manipulation creating current
inhomogeneity, not on the intrusive modification of material
properties, and is thus achievable with less technological
complication.

We emphasise that the enhancement, as a geometric effect, is
highly sensitive to the electrode locations where the signal is
probed. In Fig. \ref{fig:fig2}f, we plot the oscillations in
$R_{xx}$ measured with an additional set of \ti{outer} electrodes
V$_{xx}^{\tx{out}}$ (see Figs. \ref{fig:fig2}a), where the current
density is higher. We observe a sharp contrast to the signal
acquired at inner contacts: At outer contacts, the SdH
oscillations are buried by a larger background magnetoresistance
(compare to that measured at inner contacts), and are thus less
distinguishable.

The $R_{xx}$ data allows us to extract the effective mass
$m^{\ast}$ in the surface state by fitting the temperature
dependence of the oscillatory component $\Delta R_{xx}$ into the
Lifshits-Kosevich formula. This yields
$m^{\ast}/m_{0}=  0.0381\pm 0.0004$ which is close to $0.033\pm
0.006$ obtained by using low-temperature scanning tunneling
microscopy \cite{canali-InAs-surface-1998}.

\begin{figure}[h]
\centering
\includegraphics[scale=0.7]{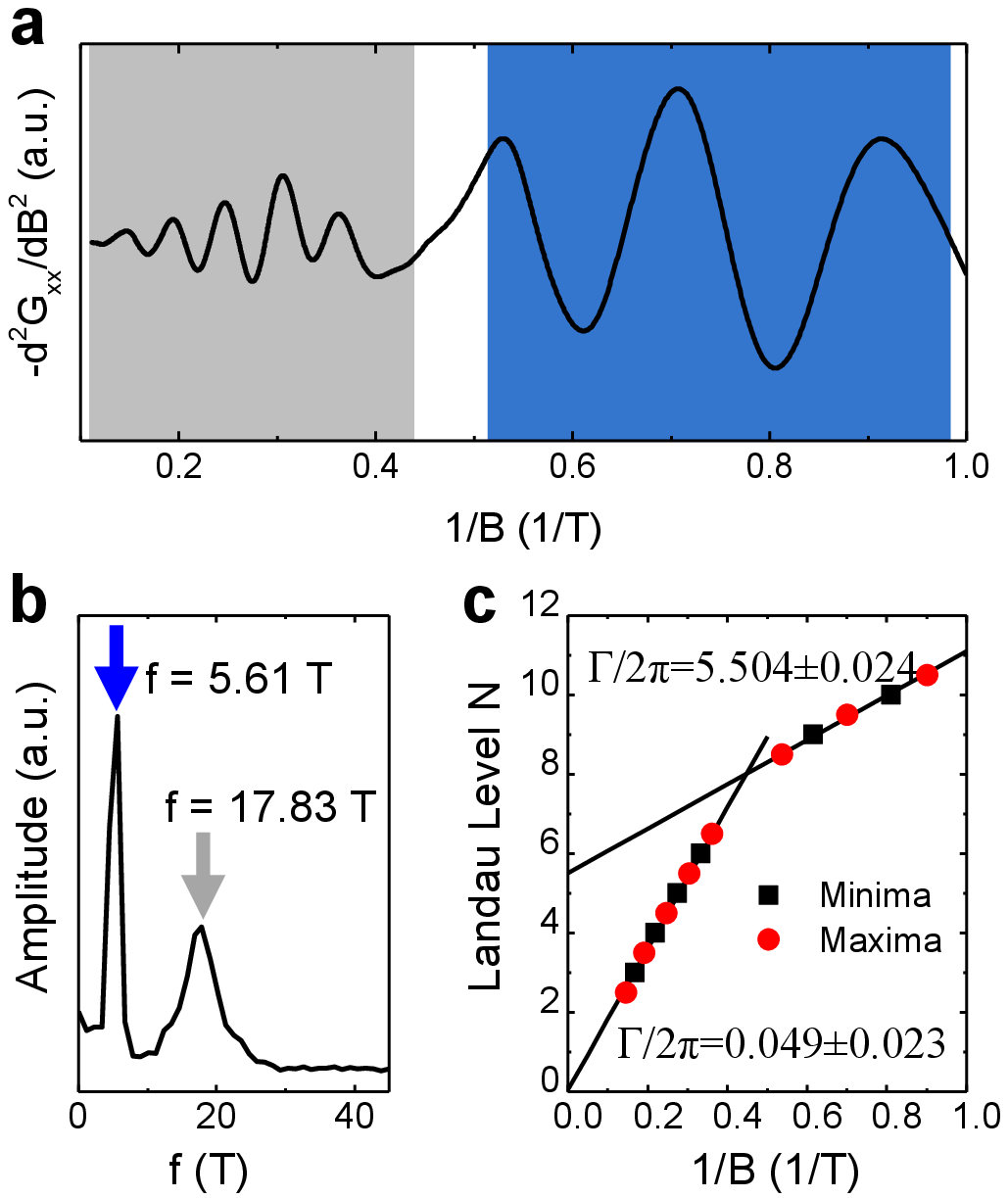}
\caption{SdH oscillations in InAs surface state (a) Oscillations in $-d^{2}G_{xx}/dB^{2}$ at perpendicular
field $\theta = 0^{\circ}$. Blue (grey) zone highlights the low
(high) field regime. (b) Fourier analysis on the
$-d^{2}G_{xx}/dB^{2}$ signal. Blue (grey) arrow indicates the
frequency at low (high) field. (c) Landau level fan diagram of
the SdH oscillations in $G_{xx}$. The
solid lines are the fittings in separate regimes using Eq.\ref{eq:berry-phase-and-field}.}
\label{fig:fig3}
\end{figure}

The frequency of the SdH oscillations also contains the
information of carrier density and Berry phase in the surface
state.
However, we notice that, in the high field
regime, the phase difference between $-d^{2}R_{xx}/dB^{2}$ and
$-d^{2}R_{xy}/dB^{2}$ is approximately $\sim 0$, instead of
$\pi/2$.
Such discrepancy is
attributed to the coexistence of both surface and bulk conducting
channels.
The conductance, not resistance, bears the correct information to
analyse the carrier density and Berry phase \cite{xiong-prb-2012,
Shrestha-prb-2014}.
Following a standard procedure\cite{xiong-prb-2012,
Shrestha-prb-2014}, we convert the resistance to conductance
and extract oscillatory extrema from
the negative second derivative $-d^{2}G_{xx}/dB^{2}$ (Fig.
\ref{fig:fig3}a). The Fourier analysis reveals a single frequency
$f_{\ssf{SdH}}^{\ssf{(L)}}=5.61$ T for the signal below $\sim 2$ T
(called low field regime thereafter, Fig.\ref{fig:fig3}b).
This
corresponds to a sheet density $\sim 1.36\times 10^{11}$
cm$^{-2}$, according to the Onsager relation
$N_{s}=(e/\pi\hbar)f_{\ssf{SdH}}$, and a Fermi wave vector $k_{F}=
1.28\times 10^{6}$ cm$^{-1}$ since $f_{\ssf{SdH}}=\hbar
k_{F}^{2}/(2e)$.
Here, $\hbar$ is Planck
constant $h$ divided by $2\pi$ and $e$ is the absolute value of
the elementary charge.
At fields higher than $\sim 2.5$ T (called high field regime
thereafter), a frequency $f_{\ssf{SdH}}^{\ssf{(H)}} = 17.83$ T is
obtained, suggesting a sheet density $\sim 4.32\times 10^{11}$
cm$^{-2}$ for a spin-filtered surface state.

The presence of two dominant frequencies in the SdH signal strongly
indicates the existence of two Fermi circles in the 2D
surface state. All experimental features are consistent with a 2D
Bychkov-Rashba model comprising a parabolic energy dispersion and
a Rashba spin-orbit coupling $H_{R}=\alp
\hat{\bm{z}}\cdot(\hat{\bm{\sg}}\times \bm{k})$. Here $\alp$ is
the Rashba coupling constant, $\bm{k}$ a 2D wave-vector, and
$\hat{\bm{\sg}}$ the Pauli matrix. To estimate
$\alp$, we employ the Onsager formula to relate the higher
frequency $f_{\ssf{SdH}}^{\ssf{(H)}}=\hbar (k_{F}^{(-)})^{2}/(2e)$
and the lower one $f_{\ssf{SdH}}^{\ssf{(L)}}=\hbar
(k_{F}^{(+)})^{2}/(2e)$ to the Fermi wave vectors $k_{F}^{(\pm)}$
that are determined by the two polarization branches originating
from the Rashba splitting, i.e.
$E_{F}=\hbar^{2}(k_{F}^{(\pm)})^{2}/(2m^{\ast})\pm\alp
k_{F}^{(\pm)}$. As the Fermi energy can be estimated using the
total carrier density $N_{s}$, we obtain in the leading order of
$\alp$,
\begin{align}
f_{\ssf{SdH}}^{\ssf{(H)}}-f_{\ssf{SdH}}^{\ssf{(L)}} \approx
2\frac{\alp m^{\ast}}{e\hbar}\sqrt{2\pi N_{s}}.
\label{eq:frequencies-and-alpha}
\end{align}
The Fermi energy is assumed to be larger than the Rashba energy
$\Dlt_{\tx{R}}\equiv \alp^{2}m^{\ast}/(2\hbar^{2})$. The total
sheet carrier density is $N_{s}\sim 5.67\times
10^{11}~\tx{cm}^{-2}$ and the frequencies are
$f_{\ssf{SdH}}^{\ssf{(H)}}=17.83~\tx{T}$ and
$f_{\ssf{SdH}}^{\ssf{(L)}}=5.61~\tx{T}$. Using Eq.
\ref{eq:frequencies-and-alpha}, we find a large Rashba constant
$\alp\sim 1.0~\tx{eV}~\tx{\AA}$. This value is comparable to
the ones in surface alloys \cite{ast-surface-alloy-prl-2007} but
still about one order of magnitude larger than those in typical semiconductor
heterojunctions \cite{nitta-rashba-2deg-prl-1997}. The
corresponding Rashba energy $\Dlt_{R}\sim 2.4~\tx{meV}$ is also in
par with the surface states in Au
\cite{lashell-au-surface-prl-1996} or Bi
\cite{koroteev-bi-prl-2004}, and quantum wells states formed on Pb
thin film \cite{dil-pb-fim-prl-2008}. The Fermi surface is about
8.54 meV from the bottom of the inner Rashba band (the band touching
point), and 27.15 meV from the bottom of the outer band.

We may also obtain, from the SdH signal, the phase shift $\gm$
encoded in the oscillating part of the longitudinal
magneto-conductance $\Dlt G_{xx}$
\begin{align}
\Dlt G_{xx}\sim
\cos\left[2\pi\left(\frac{f_{\ssf{SdH}}}{B}-\gm\right)\right]
\label{eq:cond-osc}
\end{align}
In solid systems, $\gm$ and the Berry's phase
$\Gamma$\cite{berry-phase-1984} are simply related by
\begin{align}
\gm = \frac{1}{2}-\frac{\Gamma}{2\pi}.
\label{eq:phaseshift-berryphase-simple-relation}
\end{align}
A nonzero Berry's phase, $\Gamma = \pi$ (or an integer-like
$\gm$), can be achieved at a band touching (Dirac) point
\cite{mikitik-sharlai-prl-1999} as in 2D systems like graphene
\cite{zhang-gaphene-nature-2005, novoselov-natphys-2006} or the
surface states of topological insulators
\cite{qu-science-2010,analytis-natphys-2010}. Here on the InAs
surface, the Rashba coupling provides this, too.

We now extract the phase shift $\gm$ and Berry's phase $\Gamma$ by
constructing Landau level fan diagram. Equation \ref{eq:cond-osc}
has the $N$th minimum at a magnetic field $B_{N}$ that satisfies
\begin{align}
2\pi\left(\frac{f_{\ssf{SdH}}}{B_{N}}-\gamma\right) = \pi(2N-1),
\label{eq:berry-phase-and-field}
\end{align}
which implies that, in a $N$ versus $1/B_{N}$ plot, the intercept
on the $N$ axis is $\gm$. The values of $B_{N}$ at the minima
(maxima) of $-d^{2}G_{xx}/dB^{2}$ are indexed by an integer (a
half integer), as shown in Fig. \ref{fig:fig3}a. The Landau fan
diagram in Fig. \ref{fig:fig3}c shows two well-defined regimes with
respect to the strength of magnetic field,
providing another evidence to the presence of Rashba spin-orbit
coupling\cite{eisenstein-prl-1984,stormer-prl-1983,
papadakis-spin-split-2DHG-science,tsukazaki-2007-science}.

According to Eq. \ref{eq:phaseshift-berryphase-simple-relation},
in the low field regime $B<2$ T, an intercept $\Gamma/2\pi =
5.504\pm 0.024$ is derived from a linear fitting with
$f_{\ssf{SdH}}^{\ssf{(L)}}\sim 5.61$ T obtained from the Fourier
analysis. This half-integer-like $\Gamma/2\pi$ (or an integer like
$\gm$) shows a nonzero Berry's phase $\pi$. This is a direct
evidence to the energy dispersion; the surface state possess
a $k$-linear dispersion to create a band-touching point at $k\sim
0$ \cite{mikitik-sharlai-prl-1999}. This is consistent with the
Bychkov-Rashba model \cite{bychkov-rashba-jpc-1984}. At high
field, a linear fitting suggests an integer-like $\Gamma/2\pi =
0.049\pm 0.023$ and thus a zero Berry's phase, due to the Zeeman
splitting.

By creating current inhomogeneity in an InAs Hall bar with a metal
shunt, we have achieved enhanced SdH signals. These quantum
oscillations originate from the 2D surface state with a large
Rashba coupling ($\alp\sim 1.0~\tx{eV}~\tx{\AA}$), which is
further confirmed by the observation of a nonzero Berry's phase.
We envisage that the architecture based on geometric manipulation
proposed in this letter is applicable to broader systems with bulk
and surface conduction channels. Our results even entice us to
attempt the possibility of tuning an ordinary semiconductor into a
pseudo topological insulator.

\section*{Methods}
{\bf InAs Epitaxial Sample.}  The InAs epitaxial sample is grown  by the
Veeco Gen 930 molecular beam epitaxy system on a (100) oriented semi-insulating GaAs substrate with the following structure: 1)
a 0.2 $\mu$m thick GaAs buffer layer, 2) a 1~$\mu$m-thick In$_{x}$Ga$_{1-x}$As metamorphic buffer, in
which the indium concentration is increased linearly from 0 to
1, 3) a 0.2 $\mu$m thick undoped InAs stabilizing buffer, 4) a 1.5
$\mu$m thick Si-doped active layer.
The compensation effect is negligible with a doping level of 10$^{16}$ cm$^{-3}$.

{\bf Device Fabrication.} Photo-lithographic techniques are
used for device patterning. The
InAs bar of size $40\mu\tx{m} \times 1000 \mu \tx{m}$ is wet
etched in a citric acid/H$_2$O$_2$ solution exploiting the
semi-insulating GaAs as an etch stop. The contacts are formed by
a sputtered Ti (50 nm)/Au (250 nm) stack acting as the metal shunt
and electrodes. A rapid thermal annealing process at 250 $^\circ$C
is utilized to improve the contact resistivity to 10$^{-7}
~\Omega$cm$^2$.

{\bf Device Characterization.} The magnetotransport measuremensts
are carried out using a Quantum
Design physical property measurement system (PPMS) with a
homogeneous magnetic field $B$ up to 9 T at 10 K. The devices are wire
bonded to a printed circuit board with a
four-probe measuring configuration. A current of 100 $\mu$A is
applied throughout the measurements. In the angular dependence measurements, the
device is mounted onto a sample holder, which can be rotated in
the magnetic field.

\section*{Acknowledgments}
We thank G. E. Bauer, J. Ieda and H. Adachi for fruitful discussions.


\begin{thebibliography}{99}

\bibliographystyle{Nature}

\bibitem{ko-nature-2010}
Ko, H. \textit{et al.}, Ultrathin compound semiconductor on insulator layers for high-performance nanoscale transistors. \textit{Nature} {\bf 468}, 286 (2010).
\bibitem{olsson-prl-1996}
Olsson, L. \"{O}. \textit{et al.} Charge accumulation at InAs surfaces. \textit{Phys. Rev. Lett.} {\bf 76}, 3626 (1996).
\bibitem{Kane-TI}
Hasan,M. Z. \& Kane, C. L. Colloquium: Topological insulators. \textit{Rev. Mod. Phys.} {\bf 82}, 3045 (2010).
\bibitem{shoenberg-book}
D. Shoenberg, \ti{Magnetic Oscillations in Metals} (Cambridge
Univ. Press, UK, 1984).
\bibitem{analytis-natphys-2010}
Analytis,J. G. \textit{et al.} Two-dimensional surface state in the quantum limit of a topological insulator. \textit{Nat. Phys.} {\bf 6}, 960
(2010).
\bibitem{Allgaier-JAP}
Allgaier, R. S. A new analysis of the linear high-field magnetoresistance in n-type PbTe films. \ti{J. Appl. Phys.} {\bf 59}, 1388 (1986).
\bibitem{solin-science}
Solin, S., Thio, T., Hines, D. R., \& Heremans, J. J. Enhanced room-temperature geometric magnetoresistance in inhomogeneous narrow-hap semiconductors. \textit{Science} {\bf289}, 1530 (2000).
\bibitem{meta_Lub}
Lubyshev, D. \textit{et al.} Strain relaxation and dislocation filtering in metamorphic HEMT structures grown on GaAs substrates. \ti{J. Vac. Sci. Technol. B} {\bf 19}, 1510 (2001).
\bibitem{Wieder-APL-1974}
Wieder, H. H. Transport coefficients of InAs epilayers. \textit{Appl. Phys. Lett.} {\bf 25}, 206 (1974).
\bibitem{xiong-prb-2012}
Xiong, J. \textit{et al.}, High-field Shubnikov–de Haas oscillations in
the topological insulator Bi$_2$Te$_2$Se. \textit{Phys. Rev. B} {\bf 86}, 045314 (2012).
\bibitem{qu-science-2010}
Qu, D. -X., Hor, Y. S., Xiong, J., Cava, R. J.  \& Ong, N. P.
Quantum oscillations and Hall anomaly of surface states in the topological insulator  Bi$_2$Te$_3$.
~\textit{Science} {\bf 329}, 821 (2010).
\bibitem{buttiker-prb-1988}
B\"{u}ttiker, M. Absence of backscattering in the quantum Hall effect in multiprobe conductors.
\ti{Phys. Rev. B} {\bf 38}, 9375 (1988).
\bibitem{vanwees-prb-1989}
van Wees, B. J. \textit{et al.} Suppression of Shubnikov–de Haas resistance oscillations due to selective population or detection of Landau levels: Absence of inter-Landau-level scattering on macroscopic length scales. \ti{Phys. Rev. B} {\bf 39}, 8066 (1989).
\bibitem{canali-InAs-surface-1998}
Canali, L., Wild\"{o}er, J. W. G., Kerkhof, O. \& Kouwenhoven, L. P. Low-temperature STM on InAs(110) accumulation surfaces. \textit{Appl. Phys. A} {\bf 66}, S113 (1998).
\bibitem{Shrestha-prb-2014}
Shrestha, K., Marinova, V., Lorenz, B., \& Chu, P. C. W. Shubnikov–de Haas oscillations from topological surface states of metallic Bi$_2$Se$_{2.1}$Te$_{0.9}$. \ti{Phys. Rev. B} {\bf 90}, 241111(R) (2014).
\bibitem{ast-surface-alloy-prl-2007}
Ast, C. R. \textit{et al.} Giant spin splitting through surface alloying. \textit{Phys. Rev. Lett.} {\bf 98}, 186807 (2007).
\bibitem{nitta-rashba-2deg-prl-1997}
Nitta, J., Akazaki, T., Takayanagi, H. \& Enoki, T. Gate control of spin-orbit interaction in an inverted In$_0.53$Ga$_0.47$As/In$_0.52$Al$_0.48$As heterostructure. \textit{Phys. Rev. Lett.} {\bf 78}, 1335 (1997).
\bibitem{lashell-au-surface-prl-1996}
LaShell, S., McDougall, B. A. \& Jensen, E. Spin splitting of an Au(111) surface state band observed with angle resolved photoelectron spectroscopy. \textit{Phys. Rev. Lett.} {\bf77}, 3419(1996).
\bibitem{koroteev-bi-prl-2004}
Koroteev, Yu. M. \textit{et al.} Strong spin-orbit splitting on Bi surfaces. \textit{Phys. Rev. Lett.} {\bf 93}, 046403 (2004).
\bibitem{dil-pb-fim-prl-2008}
Hugo Dil, J. \textit{et al.} Rashba-type spin-orbit splitting of
quantum well states in ultrathin Pb films. \textit{Phys. Rev. Lett.} {\bf 101}, 266802 (2008).
\bibitem{berry-phase-1984}
Berry, M. V. Quantal phase factors accompanying adiabatic changes.
\textit{Proc. R. Soc. London Ser. A} {\bf 392}, 45 (1984).
\bibitem{mikitik-sharlai-prl-1999}
Mikitik, G. P. \& Sharlai, Yu. V. Manifestation of Berry's Phase in Metal Physics. \textit{Phys. Rev. Lett.} {\bf 82},
2147 (1999).
\bibitem{zhang-gaphene-nature-2005}
Zhang, Y., Tan, Y. -W., Stormer, H. L. \& Kim, P. Experimental observation of the quantum Hall effect and Berry's phase in graphene. \textit{Nature} {\bf
438}, 201 (2005).
\bibitem{novoselov-natphys-2006}
Novoselov, K. S. \textit{et al.} Unconventional quantum Hall effect and Berry's phase of 2$\pi$ in bilayer graphene. \textit{Nat. Phys.} {\bf 2}, 177(2006).
\bibitem{eisenstein-prl-1984}
Eisenstein, J. P., St\"{o}rmer, H. L., Narayanamurti, V., Gossard, A. C. \& Wiegmann, W. Effect of inversion symmetry on the band structure of semiconductor heterostructures. \textit{Phys. Rev. Lett.} {\bf 53}, 2579 (1984).
\bibitem{stormer-prl-1983}
Stormer, H. L. \textit{et al.} Energy structure and quantized Hall effect of two-dimensional holes.
\textit{Phys. Rev. Lett.} {\bf 51}, 126 (1983).
\bibitem{papadakis-spin-split-2DHG-science}
Papadakis, S. J., De Poortere, E. P., Manoharan, H. C., Shayegan, M. \& Winkler, R. The effect of spin splitting on the metallic behavior of a two-dimensional system. \textit{Science} {\bf 283}, 2056 (1999).
\bibitem{tsukazaki-2007-science}
Tsukazaki, A. \textit{et al.} Quantum Hall effect in polar oxide heterostructures. \textit{Science} {\bf 315}, 1388 (2007).
\bibitem{bychkov-rashba-jpc-1984}
Bychkov, Yu. A. \& Rashba, E. I. Oscillatory effects and the magnetic susceptibility of carriers in inversion layers. \textit{J. Phys. C: Solid State Phys.} {\bf 17}, 6039 (1984).
\end{thebibliography}
\end{document}